\documentclass[runningheads]{llncs}
\usepackage{soul}
\usepackage{color}
\usepackage{array, makecell}
\usepackage{paralist}
\usepackage{graphicx}
\usepackage{hyperref}
\usepackage{booktabs}
\begin{document}
\title{Citation Recommendation on Scholarly Legal Articles}
%
%
\author{Doğukan Arslan\inst{1}\orcidID{0000-0002-8114-2953} \and
Saadet Sena Erdoğan\inst{2}\orcidID{0000-0001-5304-3483} \and
Gülşen Eryiğit\inst{1}\orcidID{0000-0003-4607-7305}}
\authorrunning{D. Arslan et al.}
%
\institute{Department of Artificial Intelligence and Data Engineering\\Istanbul Technical University, İstanbul, Türkiye\\
\email{\{arslan.dogukan,gulsenc\}@itu.edu.tr} \and
Department of Computer Engineering\\Istanbul Technical University, İstanbul, Türkiye\\
\email{erdogansa20@itu.edu.tr}}

\maketitle              
\begin{abstract}
Citation recommendation is the task of finding appropriate citations based on a given piece of text. The proposed datasets for this task consist mainly of several scientific fields, lacking some core ones, such as law. Furthermore, citation recommendation is used within the legal domain to identify supporting arguments, utilizing non-scholarly legal articles. In order to alleviate the limitations of existing studies, we gather the first scholarly legal dataset for the task of citation recommendation. Also, we conduct experiments with state-of-the-art models and compare their performance on this dataset. The study suggests that, while BM25 is a strong benchmark for the legal citation recommendation task, the most effective method involves implementing a two-step process that entails pre-fetching with BM25+, followed by re-ranking with SciNCL, which enhances the performance of the baseline from 0.26 to 0.30 MAP@10. Moreover, fine-tuning leads to considerable performance increases in pre-trained models, which shows the importance of including legal articles in the training data of these models.

\keywords{Citation Recommendation \and Legal Natural Language Processing \and Information Retrieval}
\end{abstract}
\section{Introduction}
The task of citation recommendation involves identifying potential citations among some candidates for a particular text, specifically for justifying arguments or making concepts clear. Predominantly, studies in this task neglect the lack of diversity and the imbalance of datasets regarding article fields, which might affect the performance of models.
The challenge of this issue has been recently addressed in \cite{Medic2022LargescaleEO}, proposing a new benchmark dataset that spans diverse scientific fields along with the field-level evaluation of various models. Nevertheless, certain core fields, such as law, remain excluded.

In the context of legal natural language processing, citation recommendation is primarily used to discover justifying arguments from non-scholarly law related articles, mostly judicial opinions \cite{Huang2021ContextawareLC,Dadgostari2020ModelingLS,Thomas2020QuickCA},
which leaves scholarly legal articles unexplored. Since it is possible to automatically generate labeled data for the task of citation recommendation, models trained on scholarly legal articles could offer considerable value for various legal natural language processing tasks, including legal case retrieval \cite{Locke2022CaseLR}, legal document similarity \cite{Bhattacharya2020MethodsFC}, and legal judgement prediction \cite{Cui2022ASO}. 

In order to address this, we introduce the very first scholarly legal citation recommendation dataset in the literature: We gather 719 scholarly legal articles with 10K incoming citation links from 9K articles. Additionally, this paper provides baseline scores for the citation recommendation task on scholarly legal articles using state-of-the-art models such as BM25 \cite{robertson1995okapi}, Law2Vec \cite{Chalkidis2018}, SciBERT~\cite{SciBERT}, SPECTER \cite{SPECTER}, LegalBERT \cite{LegalBERT}, SciNCL \cite{SciNCL}, and SGPT \cite{sgpt} in four different setups: \begin{inparaenum}[(i)]\item as a baseline, a BM25 model is trained with the gathered dataset. The performance of this model in finding relevant articles to cite is evaluated along with the aforementioned pre-trained models. \item pre-trained models are fine-tuned using the gathered dataset. Next, the task is divided into two: pre-fetching and re-ranking. Then, \item pre-trained and \item fine-tuned models are utilized for re-ranking articles retrieved by BM25. \end{inparaenum}We demonstrate that BM25 is a suitable baseline approach with competitive results for citation recommendation on scholarly legal articles. Overall, a two-step approach consisting of pre-fetching with BM25+ and subsequent re-ranking with SciNCL performs most effectively on our dataset. Besides, the performance of the fine-tuned models in comparison to the pre-trained ones, with a small-sized dataset, indicates that models trained with scholarly texts should include articles from more varied scientific fields, including the legal domain. The dataset and fully reproducible code is publicly available on GitHub\footnote{\href{https://github.com/dgknrsln/LegalCitationRecommendation}{https://github.com/dgknrsln/LegalCitationRecommendation}}.

The remainder of this paper is structured as follows. Section~\ref{related} gives a general introduction to citation recommendation methods and how they are applied to the legal domain. Section~\ref{methods} includes considered approaches. The dataset and experimental setup, along with the evaluation metrics and obtained results, are presented and discussed in Section~\ref{results}, and Section~\ref{conclusion} contains final thoughts and suggestions for future studies.

\section{Related Work}
\label{related}
Parallel to the rapid growth in scientific publication activity \cite{Bornmann2020GrowthRO}, recent papers have more outgoing citations than before \cite{Wahle2022D3AM}. This issue raises concerns about the quality of citations. Hence, the task of citation recommendation became popular. Consequently, the citation recommendation task has been studied with documents from different domains eg., patents \cite{Mahdabi2014QueryDrivenMO}, Wikipedia articles \cite{Fetahu2015AutomatedNS}, news \cite{Peng2016NewsCR} and legal cases \cite{Huang2021ContextawareLC}. In this section, we examine citation recommendation studies from a broader perspective and discuss their implications to the law field. One may refer to \cite{Frber2020CitationRA,Ma2020ARO,Medic2021ASO} for  detailed surveys regarding citation recommendation studies.

\subsection{Citation Recommendation}
Practiced methods in citation recommendation (CR) can be listed as collaborative filtering (CF), graph-based filtering (GB), and content-based filtering (CB). CF algorithms aims to match different users' preferences to recommend an article \cite{Pennock2000CollaborativeFB,McNee2002OnTR,Liu2015ContextBasedCF}. This kind of approaches can be brittle, particularly when data is sparse or when a newly included item has few or no evaluations from users (early-rater problem), or when a new user that the system has no knowledge of joins the system (cold-start problem) 
\cite{Ali2021AnOA}. In GB algorithms, a recommender system is modeled using a graph and relations between authors, papers, venues etc. \cite{Lao2010RelationalRU,Bez2011UnderstandingAS,Liang2011FindingRP}. They may encounter computational complexity issues \cite{Ali2021AnOA} and the problem of bias against old nodes in the network \cite{Ali2020AGT}. CB algorithms take advantage of papers' descriptive features (content such as title, abstract, sentences, or key-phrases) for CR \cite{Bhagavatula2018,10.1145/3312528,Amami2016}. Attaining those features is easy, and most recommendation systems rely on a CB method \cite{Beel2015}, as in this study. 

CR studies can roughly be categorized as local and global, concerning the extent of the recommendation. In  \textit{local citation recommendation} (LCR), also called \textit{context-aware citation recommendation}, the primary focus is a specific part of the input document, such as a sentence or a slightly larger window, whereas, in \textit{global citation recommendation} (GCR), the entire document \cite{Nallapati2008} or its abstract \cite{Ma2019PersonalizedSP,Mu2018QueryFocusedPC,Yang2019CitationRA} is considered. Different types of background knowledge may also be utilized in GCR studies including, but not limited to the title \cite{Bhagavatula2018,Galke2018MultiModalAA,Khadka2018UsingCT}, author information \cite{Galke2018MultiModalAA,Ma2018NewlyPS,Khadka2018UsingCT}, venue \cite{Ren2014ClusCiteEC,Yang2019CitationRA,Yu2012CitationPI}, and key-phrases \cite{Ma2019PersonalizedSP,Ma2018NewlyPS,Mu2018QueryFocusedPC}. 

Various scholarly datasets have been collected regarding training and evaluation of the CR methods. While most of the datasets consist of papers from computer science and related fields with citation links and metadata information, such as DBLP \cite{10.1145/1401890.1402008}, ACL-AAN \cite{radev-etal-2009-acl}, ACL-ARC \cite{bird-etal-2008-acl}, arXiv CS \cite{farber-etal-2018-high}, Scholarly Dataset \cite{https://doi.org/10.25540/bbch-qtt8}, and unarXiv \cite{Saier2019BibliometricEnhancedAA}, some includes papers from medicine like PubMed\footnote{\href{https://www.nlm.nih.gov/databases/download/pubmed_medline.html}{https://www.nlm.nih.gov/databases/download/pubmed\_medline.html}} and RELISH \cite{Brown2019LargeED} or from a wide variety of fields as CORE\footnote{\href{https://core.ac.uk/documentation/dataset}{https://core.ac.uk/documentation/dataset}}, S2ORC \cite{lo-wang-2020-s2orc}, CiteSeer\footnote{\href{https://csxstatic.ist.psu.edu/downloads/data}{https://csxstatic.ist.psu.edu/downloads/data}}, and MDCR \cite{Medic2022LargescaleEO}. However, none of them include articles from the field of law.

\subsection{Legal Citation Recommendation}
CR methods are applied to the legal domain, focusing mainly on finding appropriate non-scholarly legal documents to cite, such as court decisions, statutes, and law articles. One of the early works in legal CR makes use of a CF approach to build a legal recommender system \cite{Winkels2014TowardsAL}. \cite{Thomas2020QuickCA} tries to solve the legal document recommendation problem, studying it in a CR context and utilizing citation network analysis and clickstream analysis. \cite{Ostendorff2021EvaluatingDR} studies various citation-based graph methods for the legal document recommendation task, representing documents as nodes and citation links as edges in the graph.  Varying CB and CF methods are applied in \cite{Huang2021ContextawareLC} to find proper legal documents to cite. Evaluations on several metrics demonstrate the strength of neural models over traditional ones in the legal CR task. \cite{Dhani2021SimilarCR} works with legal knowledge graphs to predict citation links and find similar cases.

%

\section{Methods}
\label{methods}
The following seven approaches are utilized for the legal CR task in four different setups, which are applying pre-trained language models directly, fine-tuning pre-trained models before using, and first pre-fetching articles with BM25, then utilizing pre-trained and fine-tuned models to re-rank the retrieved results.
This multi-stage approach is often employed in information retrieval systems \cite{https://doi.org/10.48550/arxiv.1901.04085}, where faster but less accurate models like BM25 are used in the initial stage to retrieve a subset of related documents. In the subsequent stage, slower but more precise models refine the ranking of the top candidate list to improve the retrieval system's effectiveness. This separation of retrieval into stages enables the retrieval system to achieve a good trade-off between efficiency and effectiveness.

\paragraph{BM25} Given a query, Best Matching-25 scores the relevance of documents. It is a strong baseline for legal case retrieval task \cite{https://doi.org/10.48550/arxiv.2105.05686}, which is quite similar with citation recomendation. BM25+ implementation of a Python package called rank-bm25 \cite{rankbm25} is trained with the abstracts of LawArXiv articles. Then, the 10 most relevant articles are retrieved for each query article ($\sim$9K). This model is also used in the pre-fetching step as in traditional information retrieval studies.

\paragraph{Law2Vec} Law2Vec is a Word2Vec \cite{w2v} variation, trained with 123,066 legal documents including legislation, court decisions, and statues. Word2Vec attempts to place each word in a vector space such that words with similar meanings are close together by iterating over the entire corpus. Each step considers a word as well as its surroundings within a window. It is observed that Law2Vec exhibits comparable performance to the other techniques in the task of similarity measuring between legal documents \cite{Mandal2021}. There are two sets of vectors in Law2Vec: one with 100 dimensions and one with 200 dimensions. The experiments are conducted using the 200-dimensional model.

\paragraph{SciBERT} SciBERT is a BERT \cite{https://doi.org/10.48550/arxiv.1810.04805} model trained with the full text of 1.14M papers from the computer science and biomedical domain. BERT is a Transformer-based model trained for the tasks of language modeling and next-sentence prediction. Being pre-trained on a vast corpus of scientific literature, including papers from a range of fields like computer science, chemistry, and biology, SciBERT is particularly well-suited for various scientific domain tasks like text classification, named entity recognition, and question answering. SciBERT has two versions: cased and uncased. We utilized the uncased version during the experiments.

\paragraph{SPECTER} SPECTER is a Transformer-based method to generate document embeddings, specifically for scientific texts. According to the authors, it differs from other pre-trained models with its applicability to downstream tasks such as CR without any task-specific fine-tuning.

\paragraph{LegalBERT} Another applied BERT variant is called LegalBERT which is a model trained with various kinds of legal documents such as legislation, court cases, and contracts. Among the other variations, the uncased base model of LegalBERT is used for experiments.

\paragraph{SciNCL} Another BERT-based model is SciNCL, which is initialized with SciBERT weights and uses controlled sampling during training. Authors claim that SciNCL outperforms SPECTER in various metrics including CR.

\paragraph{SGPT} SGPT is a GPT-based \cite{Radford2018ImprovingLU} model, trained to obtain sentence embeddings for the task of semantic search. GPT is a powerful model for the task of next token prediction since it is trained for the purpose of language modeling. It leverages decoders to generate sentence embeddings that can be used for semantic search tasks. Through this approach, SGPT achieves superior performance compared to previous state-of-the-art sentence embedding methods.

\section{Experiments \& Discussions}
\label{results}
\subsection{Data Set \& Experimental Setup}

\label{data}
Articles used in this study are gathered from LawArXiv\footnote{\href{https://osf.io/preprints/lawarxiv}{https://osf.io/preprints/lawarxiv}}, which is an open-source scholarly legal article repository. 
It contains 1366 articles that cover 27 different legal subjects. A tool\footnote{\href{https://serpapi.com/google-scholar-api}{https://serpapi.com/google-scholar-api}} for scraping search engine results page (SERP) is used to gather citing articles from Google Scholar\footnote{\href{https://scholar.google.com/}{https://scholar.google.com/}} and more than 10K articles that cite LawArXiv articles are obtained. Then, the contents of PDF files are extracted using a Python package called pdfplumber\footnote{\href{https://github.com/jsvine/pdfplumber}{https://github.com/jsvine/pdfplumber}}. Faulty extracted or non-English articles are removed after this step. Preprocessing extracted content involves converting whole text into lowercase and removing non-ASCII characters. Abstracts of the articles are obtained by splitting the document using the keyword ``abstract''. In the end, 719 LawArXiv articles with 10,111 citation links from 8,887 citing articles are used in the experiments.

In line with other content-based global citation recommendation studies \cite{Medic2022LargescaleEO}, abstracts of the articles are used as input for the fine-tuning and inference steps for all approaches. We partitioned the dataset into separate training and testing sets, utilizing 70\% of the data for training and 30\% for testing purposes.
Models are fine-tuned for three epochs and the triplet loss function is used. The obtained document and query embeddings from the pre-trained and fine-tuned models are used to calculate cosine similarity between query-document pairs, which is used to rank documents. Sentence-Transformers \cite{reimers-2019-sentence-bert}, a framework based on Huggingface's Transformers library \cite{https://doi.org/10.48550/arxiv.1910.03771}, is used to make use of pre-trained models and fine-tuning.

\subsection{Evaluation Metrics}
Experiments' results are reported with three different metrics, which are \textit{Mean Average Precision (MAP)}, \textit{Recall}, and \textit{Mean Reciprocal Rank (MRR)}. We chose to use n=10 as the reference number for computing metrics, as on average there are 14 citation links extracted per article.

\paragraph{MAP} For a number (\textit{N}) of queries, mean average precision is the mean of the average precision (\textit{AP}) scores of each query (\textit{Q}) as in the following formula:
\[MAP = \frac{1}{N}\sum_{i=1}^{N}AP(Q_{i})\]

\paragraph{Recall} The ratio of retrieved relevant documents (\textit{True Positives}) to the total number of relevant documents (\textit{True Positives + False Negatives}) is called recall. It is calculated with the following formula:
\[Recall = \frac{TP}{TP + FN}\]

\paragraph{MRR} For a number (\textit{N}) of queries, mean reciprocal rank is the mean of the reciprocal ranks where $rank_{Q}$ points the position of the first relevant document that is retrieved for a query (\textit{Q})  as in the following formula:
\[MRR = \frac{1}{N}\sum_{i=1}^{N}\frac{1}{rank_{Q_{i}}}\]

\subsection{Results \& Discussions}
This section provides and discusses the results of our experiments under four subsections.

\begin{table}[!htp]
  \centering
  \caption{Performance of BM25 and pre-trained models for retrieving top-10 articles.}
  \label{tab:my-table}
  \begin{tabular}{lccc}
    \toprule
    &\textbf{MAP@10 ↓}&\textbf{Recall@10}&\textbf{MRR@10}\\
    \midrule
\textbf{BM25}     & \textbf{0.26}   & \textbf{0.45}      & \textbf{0.31}   \\
\textbf{SciNCL}    & 0.18           & 0.33              & 0.23            \\ 
\textbf{SGPT}      & 0.17            & 0.30               & 0.22            \\ 
\textbf{SPECTER}   & 0.14          & 0.26              & 0.19            \\
\textbf{Law2Vec}   & 0.11           & 0.21               & 0.17            \\ 
\textbf{LegalBERT} & 0.08            & 0.16               & 0.15            \\ 
\textbf{SciBERT}   & 0.08           & 0.16              & 0.14            \\ 
  \bottomrule
\end{tabular}
\end{table}

\begin{table}[!htp]
  \centering
  \caption{Performance of fine-tuned models for retrieving top-10 articles.}
  \label{tab:my-table-2}
  \begin{tabular}{lccc}
    \toprule
\textit{Fine-tuned} & \textbf{MAP@10 ↓ } & \textbf{Recall@10} & \textbf{MRR@10} \\
\midrule
\textbf{SciNCL}     & \textbf{0.26}   & \textbf{0.49}      & \textbf{0.30}   \\ 
\textbf{SciBERT}    & \textbf{0.26}           & \textbf{0.49}               & 0.29            \\ 
\textbf{SPECTER}    & 0.25            & 0.47               & 0.28            \\ 
\textbf{SGPT}       & 0.25            & 0.46               & 0.29            \\ 
\textbf{LegalBERT}  & 0.24            & 0.47               & 0.28            \\ 
  \bottomrule
\end{tabular}
\end{table}

\begin{table}[!htp]
  \centering
  \caption{Performance of pre-trained models for re-ranking top-10 articles retrieved by BM25.}
  \label{tab:my-table-3}
  \begin{tabular}{lccc}
    \toprule
\textit{BM25 Prefetch} & \textbf{MAP@10 ↓ } & \textbf{Recall@10} & \textbf{MRR@10} \\ \midrule
\textbf{SciNCL}           & \textbf{0.25}            & \textbf{0.45}               & \textbf{0.30}            \\ 
\textbf{SGPT}             & \textbf{0.25}   & \textbf{0.45}      & \textbf{0.30}   \\ 
\textbf{SPECTER}          & 0.24            & \textbf{0.45}               & 0.29            \\
\textbf{LegalBERT}        & 0.19            & \textbf{0.45}               & 0.24            \\ 
\textbf{Law2Vec}          & 0.19            & \textbf{0.45}               & 0.24            \\
\textbf{SciBERT}          & 0.19            & \textbf{0.45}               & 0.24            \\ 

  \bottomrule
\end{tabular}
\end{table}

\begin{table}[!htp]
  \centering
  \caption{Performance of fine-tuned models for re-ranking top-10 articles retrieved by BM25.}
  \label{tab:my-table-4}
  \begin{tabular}{lccc}
    \toprule
\makecell{\textit{BM25 Prefetch} \\\textit{+ Fine-tuning}}
 & \textbf{MAP@10 ↓ } & \textbf{Recall@10} & \textbf{MRR@10} \\ \midrule
 \textbf{SciNCL}                                                                   & \textbf{0.30}            & \textbf{0.45}               & \textbf{0.34}            \\ 
 \textbf{SGPT}                                                                     & 0.29            & \textbf{0.45}               & \textbf{0.34}           \\
 \textbf{SciBERT}                                                                  & 0.29            & \textbf{0.45}               & \textbf{0.34}            \\ 
\textbf{SPECTER}                                                                  & 0.28   & \textbf{0.45}      & 0.33   \\ 
\textbf{LegalBERT}                                                                & 0.28            & \textbf{0.45}               & 0.33            \\ 
  \bottomrule
\end{tabular}
\end{table}

\paragraph{BM25 and Pre-trained Models:} Comparison of BM25 and pre-trained models (Table~\ref{tab:my-table}) shows that SciBERT, which is trained with scientific documents, has no understanding of legal texts. On the other hand, the poor performance of Law2Vec and LegalBERT, which are language models specific to the legal domain, might be explained as they were not trained for the task of CR. Observable performance increase (from 0.08  to 0.24 MAP@10) of LegalBERT after fine-tuning for the CR task (Table~\ref{tab:my-table-2}) also supports this claim.  The pre-trained  SGPT (Table~\ref{tab:my-table}) outperforms other pre-trained models, except SciNCL, even though it is not trained on scientific texts. This occurs possibly because it is directly trained for semantic search which is a task highly related to CR. Hence, it is more successful at retrieving the top-$k$ articles for a given query. Yet, BM25 exceeds the performance of all pre-trained models and shows that it is a strong baseline (0.26 MAP@10) for the legal CR task, performing on par with existing performances in the literature \cite{Medic2022LargescaleEO} for other domains.

\paragraph{Fine-tuned Models:} When the pre-trained language models are fine-tuned (Table~\ref{tab:my-table-2}) using the gathered dataset for the task of CR, SciNCL and SciBERT performs relatively better than others, in accordance with  claims of~\cite{SciNCL}. Performance increases on all models show that when domain knowledge is provided, those models are also able to adapt to the legal domain. Besides, the significant improvement in LegalBERT's performance suggests that the model can effectively leverage domain-specific knowledge once it is trained on the task at hand. 

\paragraph{Pre-fetching with BM25 and Re-ranking with Pre-trained Models:} Parallel with the results of the first experiment, SciNCL and SGPT (0.25 MAP@10) stand out as the best-performing pre-trained models in re-ranking BM25's pre-fetched articles (Table~\ref{tab:my-table-3}). While the performance of all models is increased, they still do not reach the level of our baseline (BM25).

\paragraph{Pre-fetching with BM25 and Re-ranking with Fine-tuned Models:} Table~\ref{tab:my-table-4} shows the performance of fine-tuned models on re-ranking BM25's pre-fetched articles. All fine-tuned models increase the performance of BM25, and demonstrate  greater success than the second experiment where there is no pre-fetching step. Overall, SciNCL stands out as the best performing model among others, improving performance of BM25 (from 0.26 to 0.30 MAP@10).  

\section{Conclusion}
\label{conclusion}
Our study presents the first scholarly legal citation recommendation dataset in the literature, consisting of 719 scholarly legal articles with 10K incoming citations from 9K articles, to make up for the lack of scholarly legal articles in the scientific text datasets. Using the gathered dataset, experimental results with state-of-the-art models are reported in four different setups. In conclusion, our findings indicate that BM25 serves as a strong baseline for citation recommendation on scholarly legal articles, while combination of BM25 and SciNCL for pre-fetching and re-ranking, respectively, produces the most effective result on our dataset, increasing the performance of the baseline (BM25) from 0.26 to 0.30 MAP@10. It is clear that pre-trained language models perform well on this task when further trained with a suitable dataset, on which future research should focus. The size of the gathered dataset is small when compared to the other domains used in the citation recommendation literature. This might affect the performance of the models in learning the task and domain, even though the entire LawArXive collection is gathered, as stated in Section~\ref{results}. As a future work, one may think of enlarging this dataset with articles from law journals not indexed within LawArXiv.

\section*{Acknowledgement}
The numerical computations reported in this article are performed via TUBITAK ULAKBIM High Performance and Grid Computing Center.

%
%
%
\bibliographystyle{splncs04}
\bibliography{sample-base}

\end{document}